\documentclass[10pt,letterpaper]{article}
\usepackage{opex3}
\usepackage{graphicx}
\usepackage{dcolumn}
\usepackage{bm}
\usepackage{array}
\usepackage{hyperref}
\usepackage{amsmath}
\usepackage{multirow}
\usepackage{cite}

\newcommand{\beq}{\begin{equation}}
\newcommand{\beql}[1]{\begin{equation}\label{#1}}

\newcommand{\eeq}{\end{equation}}
\newcommand{\bsp}{\begin{split}}
\newcommand{\esp}{\end{split}}
\newcommand{\ket}[1]{\left\vert #1 \right\rangle}

\newcommand{\braa}{\left\langle}
\newcommand{\kett}{\right\rangle}

\newcommand{\bra}[1]{\left\langle #1 \right\vert}

\newcommand{\Eq}[1]{Eq.~(\ref{#1})}
\newcommand{\Equation}[1]{Equation~(\ref{#1})}

\begin{document}
\title{General coupled mode theory in non-Hermitian waveguides}
\author{Jing Xu$^{1,2}$ and Yuntian Chen$^{1,2*}$}
\address{School of Optical and Electronic Information, Huazhong University of Science and Technology, Wuhan, China.}
\address{Wuhan National Laboratory of Optoelectronics, Huazhong University of Science and Technology, Wuhan, China.}
\email{$^*$yuntian@hust.edu.cn}

\begin{abstract}
In the presence of loss and gain, the coupled mode equation on describing  the mode hybridization of various waveguides or cavities, or cavities coupled to waveguides becomes intrinsically non-Hermitian. In such non-Hermitian waveguides, the standard coupled mode theory fails. We generalize the coupled mode theory with a properly defined inner product based on reaction conservation. We apply our theory to the non-Hermitian parity-time symmetric waveguides, and obtain excellent agreement with results obtained by finite element fullwave simulations. The theory presented here is typically formulated in space to study coupling between waveguides, which can be  transformed into time domain by proper reformulation to study coupling between non-Hermitian resonators.  Our theory has the strength of studying  non-Hermitian optical systems with inclusion of  the full vector fields, thus is useful to study and design  non-Hermitian devices that support asymmetric and even nonreciprocal light propagations.
\end{abstract}

\ocis{(240.6680) bianisotropic medium; (230.7370) chirowaveguides; (230.6080) metamaterials.}

\section{Introduction}

Coupled mode theory (CMT) can be traced back to 3 decades ago \cite{Schlkunoff}, and has been an indispensable tool of analyzing and designing waveguides, resonators, couplers or many other devices from microwave frequency \cite{Haus1958,Louisell1960,Marcuse1971} to optical frequency \cite{SnyderJosa1972,YarivJQE1973,HardyJLT1985,HausJLT1987,ChuangJLT1987,HuangJOSAA1994,FanPRB1999,XuPRE2000,FanJOSAB2003}, in both time and space domain. CMT has its mathematical root of variational principle \cite{Berk1955}, which yields a stationary value for target physical quantities of the coupled optical systems, such as propagation constant, eigen-frequencies, or impedance and so on. CMT was invented  to study parametric amplifiers, oscillators, and frequency converters in microwaves \cite{Louisell1960}, with a rigorous derivation given by Schelkunoff  using mode expansion \cite{Schlkunoff}, and Haus using a variational principle \cite{Haus1958}. The CMT for optical waveguides was developed by many authors \cite{Marcuse1971,SnyderJosa1972,YarivJQE1973}, and  further refined by Hardy et al., using non-orthogonal coupled mode formulation (NCMT) \cite{HardyJLT1985,HausJLT1987,ChuangJLT1987}, in contrast to the orthogonal coupled mode theory (OCMT). Subsequently, temporal coupled mode theory was applied to study coupling among optical resonators, or resonators coupled to waveguides \cite{FanJOSAB2003,FanPRB1999,XuPRE2000}.

It can be proved that CMT is exactly equivalent to Maxwell's equations, as long as a complete set of modes is taken in the mode expansion in constructing the coupled mode equations. In most cases, we only need a few modes (usually two) in the mode expansion, because all the other modes, for instance, the continuum modes in waveguides, do not couple to the target mode that we are interested in. Even though CMT remains approximate in the truncated mode set, yet insightful and often accurate description of target quantities of a coupled system can be obtained. By excluding those irrelevant modes, CMT for a two-modes structure can be reduced into a form of 2$\times$2 matrix, which is a classic model and has been extensively used in describing many different coupled physical systems, such as two coupled mechanical oscillators, light (single-mode ) matter (single atom) interaction in Jaynes-Cummings model, two coupled resonators in electric circuits and so on. Mostly, for a lossless system, the energy flow from one mode/oscillator/atom to another mode/oscillator/photon is reciprocal, namely, the energy will flow back in a 'reciprocal' way as 'time' is reversed, which renders the coupled system Hermitian, or the transition matrix of the coupled system as a Hermitian Matrix.

The aforementioned formulations of CMT rely on a definite and conserved optical power of the whole coupled system for either OCMT or NCMT, which essentially by default selects a scheme of complex inner product. However,  for coupled optical parity-time ($\mathcal{PT}$)-symmetric structures \cite{EIganainyOL2007,PRL2008Moiseyev,Lin2011,Feng2013,Zhu2014PRX,Fan2015,Ruter2010,Hodaei2014,AlaeianPRB2014,Dionne2014pra,Chenlin2014} the total integrated power is not a conserved quantity, especially in the broken phase after the exceptional point. Therefore, the standard CMT fails in such non-Hermitian systems \cite{EIganainyOL2007}. In \cite{EIganainyOL2007}, the authors formulate a CMT to studying $\mathcal{PT}$-symmetric structures through a Lagrangian treatment, in which the vector optical field is approximate as a  scalar field envelope. In this work, we provide a conceptually simple model to construct general coupled mode theory  (GCMT) using reaction conservation, a well-known concept in electromagnetism. The GCMT is capable of capturing the vectorial nature of the optical fields. Particularly, using a scalar inner product, we prove that the Maxwell equation remains self-adjoint, even in the presence of loss and gain. We further construct GCMT  based on perturbation. In the application of GCMT, we study $\mathcal{PT}$-symmetric waveguides with balanced losses and gain. The prediction of our theory shows excellent agreement with results obtained by the fullwave simulation (COMSOL).

The paper is organized as follows. In Section 2, we  give the foundation of our general coupled theory based on reaction concept. Secondly, we discuss the procedures of constructing GCMT. In Section 3, we study the mode dispersion of $\mathcal{PT}$-symmetric waveguides using GCMT, and compare it with conventional coupled mode theory. Finally, Section 4 concludes the paper.

\section{General couple mode theory}

\subsection{Reaction concept and self-adjointness}
The reaction  is a physical observable introduced by  Rumsey \cite{Rumsey1954} to  measure the reaction between sources $\bm S_a$ and $\bm S_b$, defined as follows,
\beq\label{reaction}
(\bm F_b, \bm S_a)=\braa  \bm F_b , \sigma \bm S_a  \kett
\eeq
where $\bm F_b$ is the field generated by source $\bm S_b$. As seen from \Eq{reaction},  the inner-product $(\cdot | \cdot)$ is defined for the vector field, and equals  $\braa \cdot \sigma \cdot  \kett$, where $\braa \bm \phi(\bm r) | \bm \psi(\bm r)  \kett = \int dr \bm \phi^T(\bm r) \bm \psi(\bm r) $ for two column vector fields  $\bm \phi(\bm r)$, $\bm \psi(\bm r)$. It is important to point out that there is no complex conjugate operation over the fields, therefore the relation $\braa \bm \phi(\bm r) | \bm \psi(\bm r)  \kett =\braa \bm \psi(\bm r) | \bm \phi(\bm r)  \kett $ holds. Denoting the field $\bm F=[\bm E, \bm H]^T$, and  $\bm S=[\bm J, \bm M]^T$, $\bm J$ ($\bm M$) electric (magnetic) current density. The reaction given in Eq.~(\ref{reaction}) can be explicitly given by
\beq
(\bm F_b, \bm S_a)=\int dr
\left[
\begin{array}{cc}
\bm E_b  & \bm H_b
\end{array}
\right] \sigma  \left[
\begin{array}{c}
\bm J_a  \\
\bm M_a
\end{array}
\right]
\eeq
where the metric tensor $\sigma=\left(
\begin{array}{cc}
\bar{\bm{1}}  & \bar{\bm{0}}  \\
\bar{\bm{0}}  & -\bar{\bm{1}}
\end{array}
\right)$,  $\bar{\bm{1}}$ denotes the identity matrix, $\bar{\bm{0}}$ the zero matrix. The link between the source $\bm S$ and the associated vector fields $\bm F$ is given by the Maxwell's equations,
\beq\label{Max}
\bar{\bm{L}} \bm F=\bar{\bm{W}}\bm{S},
\eeq
where $\bar{\bm{L}} =\left(
\begin{array}{cc}
\nabla  \times   & i k_0  \bar{\bm{\mu}}_r \\
-i k_0 \bar{\bm{\epsilon}}_r & \nabla  \times
\end{array}
\right)$, $\bar{\bm{W}}=\left(
\begin{array}{cc}
\bar{\bm{0}}  & -\bar{\bm{1}}  \\
\bar{\bm{1}}  & \bar{\bm{0}}
\end{array}
\right)$.

Based on the reaction concept, the reciprocity theorem can be given
\beq\label{reciprocity}
(\bm F_a, \bm S_b)=(\bm S_a, \bm F_b),
\eeq
which states that the response of  one source to the external field induced by another source equals to the response of the second source to the field given by the first source. \Equation{reciprocity} imposes certain constraints on the material parameters as given by $\bar{ \bm{\epsilon}}_r(\bm r)= \bar{ \bm{\epsilon}}_r^T(\bm r)$ and $\bar{\bm{\mu}}_r(\bm r)= \bar{\bm{\mu}}_r^T(\bm r)$. Such  medium is called reciprocal medium in the literature, and could be lossy or active, as long as reciprocal conditions are fulfilled. We can reformulate \Eq{Max} in a matrix form as follows,
\beq\label{Max2}
\bar{\bm{H}} \bm F=\bm{S},
\eeq
where $\bar{\bm{H}}=\bar{\bm{W}}^{-1} \bar{\bm{L}}$. For any reciprocal medium, we find that for any two vector fields $\bm \psi$, $\bm \phi$, the following relation holds:
\beq\label{SA}
(\bm \psi,\bar{\bm{H}} \bm \phi)=(\bar{\bm{H}}\bm \psi, \bm \phi).
\eeq
\Equation{SA} means that the operator $\bar{\bm{H}}$ is self-adjoint in the scheme of the inner-product given by \Eq{reaction}, which has relevant consequences and implications that we intend to discuss. Firstly, the self-adjointness of the operator $\bar{\bm{H}}$  and the reaction for defining  reciprocity share the same definition on the  inner-product. Moreover, the self-adjointness of $\bar{\bm{H}}$  is equivalent to reciprocity theorem. In other words, the reciprocal conditions on the material parameters are necessary and sufficient condition both  to   self-adjointness of $\bar{\bm{H}}$, and to reciprocity theorem. Secondly, the self-adjointness of the Hamiltonian $\bar{\bm{H}}$ guarantees that the underlying space associated with the operator $\bar{\bm{H}}$ is complete, despite the existence of losses or gain in the material parameters. Lastly, the left eigenvector space (bra space $\bra \cdot$) and the right eigenvector space (ket space $\ket \cdot$)  are closely related with each other for self-adjoint operator \cite{Friedman1962}, namely, one determines the other, and vice versa. For complex inner-product, the self-adjoint operator is conventionally called as Hermitian operator. The matrix form of right and left eigenvectors  are complex conjugate and transpose to each other, i.e., $\ket \cdot=[(\bra \cdot)^T]^*$. As for a scalar inner-product, the matrix form of self-adjoint operator is symmetric, hence the  left and right eigenvectors has the relation of  $\ket \cdot=(\bra \cdot)^T$.

\subsection{Dimension reduction: a non self-adjoint formulation for waveguide problems based on variational principles}
It is important to realize that the self-adjointness of the Maxwell's equations in reciprocal medium, as given in \Eq{SA}, is valid for the inner-product defined in 3D space. As for the development of GCMT or the modal solver based on variational principle for   waveguide problems, it is necessary to reduce the 3D formulation into its 2D counterpart.

Considering an infinitely long waveguide, the waveguide modes of the original problem are given by $\bm \phi=[\bm{e}(\bm{r}),  \bm{h}(\bm{r})]^T$, where $\bm{e}(\bm{r})=\bm{e}(x,y, \beta )e^{i(\omega t-\beta z)}$, $\bm{h}(\bm{r})=\bm{h}(x,y,\beta)e^{i(\omega t-\beta z)}$. The adjoint fields are given by  $\bm \psi=[\bm{e}^a(\bm{r}),  \bm{h}^a(\bm{r})]^T$, where $\bm{e}^a(\bm{r})=\bm{e}^a(x,y)e^{i(\omega t+\beta z)}$, $\bm{h}^a(\bm{r})=\bm{h}^a(x,y)e^{i(\omega t +\beta z)}$. We choose  counter-propagating modes  with the same $\beta$ as our mode sets $\bm \phi =[\bm{e}^+,  \bm{h}^+]^T e^{i(\omega t-\beta z)}$, and $\bm \psi=[\bm{e}^-,  \bm{h}^-]^T e^{i(\omega t+\beta z)}$ respectively.  The relation between the fields propagating in $+z$ and $-z$ direction is given by $\bm{e}^+=\{e_x,e_y,e_z\}$, $\bm{h}^+=\{h_x,h_y,h_z\}$, $\bm{e}^-=\{e_x,e_y,-e_z\}$, $\bm{h}^-=\{-h_x,-h_y,h_z\}$.
The particular choice of the mode sets \cite{SiegmanLasersBook,ChenLien1980,Pintus2014OE,Harrington2001} is relevant: (1) it is necessary to get a functional of coupled modes that is independent of $z$, meaning the terms $e^{\pm i\beta z}$ in the inner-product between the modes sets $\bm \phi$  and $\bm \psi$ are canceled;  (2) the mode pair of counter-propagating modes has a definite relation, hence $\bm \phi$ can be deduced from $\bm \psi$, and vice versa.  It is easy to get the $z$-independent and source-free wave-equation for the mode profiles $\bm \phi_{2d}=[\bm{e}^+,\bm{h}^+]$ of the original problem as follows
\beq\label{H2d}
\bar{\bm{H}}_{2d} \bm{\phi}_{2d}=0
\eeq
where  $\bar{\bm{H}}_{2d}=\left(
\begin{array}{cc}
\nabla_t \times - i\beta \bm{z}\times  & i k_0  \bar{\bm{\mu}}_r \\
-i k_0 \bar{\bm{\epsilon}}_r & \nabla_t \times - i\beta \bm{z}\times
\end{array}
\right) $, and $\bar{ \bm{\epsilon}}_{r}^a=\left(
\begin{array}{ccc}
 \epsilon ^{tt}_{r} & \epsilon ^{tz}_{r} \\
\epsilon ^{zt}_{r} & \epsilon ^{zz}_{r}
\end{array}
\right)$, and $\bar{ \bm{\mu}}_{r}^a=\left(
\begin{array}{cc}
 \mu ^{tt}_{r} & \mu ^{tz}_{r}  \\
\mu ^{zt}_{r} & \mu ^{zz}_{r}  \\
\end{array}
\right)$, and $\epsilon ^{tt}_{r}$ ($\mu ^{tt}_{r}$) denotes the $2\times 2$ in-plane components of electric (magnetic) dielectric function, $\nabla_t=\bm{x}\frac{\partial}{\partial x}+\bm{y}\frac{\partial}{\partial y}$. From \Eq{H2d}, and the predefined modes of $\bm \phi$ and $\bm \psi$, one can also obtain the adjoint system from the original ones ($\bar{\bm{H}}_{2d}$), as given by
\beq\label{H2da}
\bar{\bm{H}}_{2d}^a \bm{\psi}_{2d}=0
\eeq
where $\bm{\psi}_{2d}=[\bm{e}^-,\bm{h}^-]$, $\bar{\bm{H}}^a_{2d}=\left(
\begin{array}{cc}
\nabla_t \times + i\beta \bm{z}\times   & i k_0  \bar{\bm{\mu}}_r^a \\
-i k_0 \bar{\bm{\epsilon}}_r^a & \nabla_t \times + i\beta \bm{z}\times
\end{array}
\right)$, $\bar{ \bm{\epsilon}}_{r}^a=\left(
\begin{array}{ccc}
 \epsilon ^{tt}_{r} & -\epsilon ^{tz}_{r} \\
-\epsilon ^{zt}_{r} & \epsilon ^{zz}_{r}
\end{array}
\right)$, and $\bar{ \bm{\mu}}_{r}^a=\left(
\begin{array}{cc}
 \mu ^{tt}_{r} & -\mu ^{tz}_{r}  \\
-\mu ^{zt}_{r} & \mu ^{zz}_{r}  \\
\end{array}
\right)$. It is easy  to find that the following relation holds
\beq\label{SA2d}
(\bm \psi_{2d},\bar{\bm{H}}_{2d} \bm \phi_{2d})=0=(\bar{\bm{H}}_{2d}^a\bm \psi_{2d}, \bm \phi_{2d}),
\eeq
where the inner-product is carried out over 2D computational domain, i.e., transverse plane of the waveguides. As regards to the comparison between 2D formula (\Eq{SA2d}) and 3D formula (\Eq{SA}), a few remarks may deserve attentions. Firstly, the operator $\bm H_{2d}$ with inner product defined over 2D domain  is not self-adjoint \cite{Pintus2014OE,Harrington2001,ZhuJLT2011}, e.g., $\bar{\bm{H}}_{2d}^a\neq\bar{\bm{H}}_{2d}$, for reciprocal medium.  Secondly, for self-adjoint system, the adjoint system and the original system can be treated separately.  As for a non self-adjoint electromagnetic problem \cite{ChenLien1980}, one need to solve original problem, as well as its adjoint problem simultaneously to provide a complete but biorthogonal mode sets to construct the coupled mode equations, or any other modal solver based on variational principles, e.g., method of moments (MoM) and finite element method (FEM). As for \Eq{SA2d}, it is necessary to obtain a unified variational form that contains contribution from both  $\bm H_{2d}^a$  and  $\bm H_{2d}$. The explicit  unified variational form for eigen-mode problem of  waveguides is the following,
\beq\label{L_not_sa}
Y=(\delta\bm \psi_{2d},\bar{\bm{H}}_{2d} \bm \phi_{2d})+(\bar{\bm{H}}_{2d}^a\bm \psi_{2d}, \delta\bm \phi_{2d})=0.
\eeq
which is the first variation, e.g.,  $\delta\bm\psi_{2d}$ and $\delta\bm\phi_{2d}$, to the functional $I=(\bm \psi_{2d},\bar{\bm{H}}_{2d} \bm \phi_{2d})$. According to variational principle, \Eq{L_not_sa}  indicates simultaneously the optimal solution of $\bm\psi_{2d}$ and $\bm \phi_{2d}$ to \Eq{H2d} and \Eq{H2da}, as long as the functional  $I$ is stationary for any $\bm\psi_{2d}$ and $\bm\phi_{2d}$.  Thirdly, the mode set ($\bm \psi_{2d}$) of the adjoint system is selected as the countering propagating modes of the original system ($\bm \phi_{2d}$). As such, the total degree of freedom of unknows  is halved \cite{Pintus2014OE}. We note that GCMT can also be applied to bianisotropic waveguides \cite{Jingbia1}.

\subsection{Procedures of constructing GCMT for waveguide problem based on perturbation}
Given a perturbation  to the adjoint system  $\bar{\bm{H}}_{2d}^a$ by $\Delta\bar{\bm{H}}$, i.e.,  $\bar{\bm{H}}^{\#}_{2d}=\bar{\bm{H}}_{2d}^a + \Delta\bar{\bm{H}} $, we shall have the  relation according to linear response of Maxwell's equations, $(\bm \psi_{2d},\bar{\bm{H}}_{2d} \bm \phi_{2d})=([\bar{\bm{H}}^{\#}- \Delta\bar{\bm{H}}]\bm \psi_{2d}, \bm \phi_{2d})$, which can  be  reformulated as
\beq
(\bm \psi_{2d},\bar{\bm{H}}_{2d} \bm \phi_{2d})-(\bar{\bm{H}}^{\#}_{2d}\bm \psi_{2d}, \bm \phi_{2d})=(- \Delta\bar{\bm{H}}\bm \psi_{2d}, \bm \phi_{2d}).
\eeq
Perturbation implies that $\Delta\bar{\bm{H}}$ is small, hence $(- \Delta\bar{\bm{H}}\bm \psi, \bm \phi)$ can be approximately taken as 0, which leads to
\beq\label{MasterEquation}
(\bm \psi_{2d},\bar{\bm{H}}_{2d} \bm \phi_{2d})-(\bar{\bm{H}}^{\#}_{2d}\bm \psi_{2d}, \bm \phi_{2d})=0.
\eeq
\Equation{MasterEquation}  gives the connection between the original system and the perturbed adjoint system via the reaction conversation under a small perturbation $\Delta\bar{\bm{H}}$, and can be transcribed into a set of coupled mode equations, which is essentially the GCMT proposed in this paper. Firstly, we use normalized  fields $\bm{e}(\bm{r})=\bm{e}^+e^{i(\omega t-\beta z)}$, $\bm{h}(\bm{r})=\bm{h}^+e^{i(\omega t-\beta z)}$ propagating in the $+z$ direction, satisfying Maxwell equations
\begin{subequations} \label{Maxwellwaveguide1}
\begin{align}
\nabla_t\times\bm{e}^{+}_{0,i}-i\beta_{0,i}\bm{z}\times\bm{e}^{+}_{0,i}=-ik_0\bar{\bm{\mu}}_r^0\bm{h}^{+}_{0,i},\\
\nabla_t\times\bm{h}^{+}_{0,i}-i\beta_{0,i}\bm{z}\times\bm{h}^{+}_{0,i}=ik_0\bar{\bm{\epsilon}}_r^0\bm{e}^{+}_{0,i},
\end{align}
\end{subequations}
where the subscripts  $0$ and $i$ stand for no perturbation case and mode labels, respectively.

We consider there is a small perturbation of $\epsilon$. The idea is that the fields under perturbed system can be approximated by a linear combination of the unperturbated fields of the adjoint systems.  Therefore, the fields of perturbed system could be written as $\bm e'(\bm r)=(\Sigma_{j} a_{j} \bm{e}^{-}_{0,j} )e ^{i(\omega t+\beta z)}$, $\bm h'(\bm r)=(\Sigma_{j} a_{j} \bm{h}^{-}_{0,j}) e^{i(\omega t+\beta z)}$ and satisfy
\begin{subequations} \label{waveguide2}
\begin{align}
 \nabla_t\times\Sigma_{j} a_{j} \bm{e}^{-}_{0,j}+i\beta\bm{z}\times\Sigma_{j} a_{j} \bm{e}^{-}_{0,j}=-ik_0\bar{\bm{\mu}}_r\Sigma_{j} a_{j} \bm{h}^{-}_{0,j},\\
\nabla_t\times \Sigma_{j} a_{j} \bm{h}^{-}_{0,j} +i\beta\bm{z}\times\Sigma_{j} a_{j} \bm{h}^{-}_{0,j} =ik_0\bar{\bm{\epsilon}}_r\Sigma_{j} a_{j} \bm{e}^{-}_{0,j},
\end{align}
\end{subequations}

In equivalence with \Eq{MasterEquation}, we derive GCMT from perturbation as follows,
\begin{equation}\label{foureqadd}
\iint \{Eq.~(\ref{Maxwellwaveguide1}a)\cdot\Sigma_{j} a_{j} \bm{h}^{-}_{0,j}-Eq.~(\ref{waveguide2}b)\cdot\bm{e}^{+}_{0,i}+Eq.~(\ref{Maxwellwaveguide1}b)\cdot\Sigma_{j} a_{j} \bm{e}^{-}_{0,j}-Eq.~(\ref{waveguide2}a)\cdot\bm{h}^{+}_{0,i}\}dxdy
\end{equation}
In case that a small perturbation is present in the imaginary part of $\bar{\bm{\epsilon}}_{r}$, i.e. $\bar{\bm{\epsilon}}_{r}=\bar{\bm{\epsilon}}_{r}^{0}+i\Delta\epsilon(x,y)$, the formula resulted from  \Eq{foureqadd} can be simplified as follows,
\begin{equation}\label{cmt_bia}
\begin{split}
& \Sigma_{j} a_{j} [k_{ij}+b_{ij} -i ( \beta-\beta_{0,i})p_{ij}]= 0
\end{split}
\end{equation}
where $b_{ij}=\iint \{\nabla_t \cdot  (  \bm{h}^{+}_{0,i} \times \bm{e}^{-}_{0,j} )- \nabla_t \cdot  (  \bm{h}^{-}_{0,j} \times \bm{e}^{+}_{0,i} )\} dxdy$, $p_{ij}=\iint \{ \bm{z} \cdot ( \bm{e}^{-}_{0,j}  \times \bm{h}^{+}_{0,i})- \bm{z} \cdot ( \bm{e}^{+}_{0,i}  \times \bm{h}^{-}_{0,j})\}dxdy $, $k_{ij}=ik_0\iint \Delta\epsilon(x,y)  \bm{e}^{-}_{0,j} \cdot \bm{e}^{+}_{0,i}  dxdy$. For $\Delta\epsilon(x,y)=0$,  we shall have the following relation
\beq\label{bij}
b_{ij} = i\left( {{\beta _{0,j}} - {\beta _{0,i}}} \right){p_{ij}}.
\eeq
Inserting $b_{ij}$ back into Eq.~(\ref{cmt_bia}) yields
\begin{equation}\label{gcmt}
\begin{split}
&  \Sigma_{j} a_{j}   ( \beta -\beta_{0,j}) p_{ij}=  \Sigma_{j} a_{j} k_{ij}.
\end{split}
\end{equation}
\Equation{gcmt} is the matrix form of  GCMT proposed in this paper,  which will be used to study Hermitian  and non-Hermitian waveguide with some concrete examples in the following section.

It is worthy to write down the CMT derived from conventional CMT (CCMT)\cite{HausJLT1987} for comparison. To this end, Eq.~(\ref{MasterEquation}) shall be reformulated as
\begin{equation}
(\bm \psi^*_{2d},\bar{\bm{H}} \bm \phi_{2d})-((\bar{\bm{H}}^{\#}\bm \psi_{2d})^*, \bm \phi_{2d})=0,
\end{equation}
where $*$ indicates the operation of complex conjugation, and the metric tensor is modified as $\sigma=\left(
\begin{array}{cc}
\bar{\bm{1}}  & \bar{\bm{0}}  \\
\bar{\bm{0}}  & \bar{\bm{1}}
\end{array}
\right)$ accordingly. Following the same procedure, we have
\begin{equation}\label{gcmtconj}
\begin{split}
&  \Sigma_{j} a_{j}   ( \beta^* -\beta^*_{0,j}) p_{ij}=  \Sigma_{j} a_{j} k_{ij}.
\end{split}
\end{equation}
where $b_{ij}=\iint \{\nabla_t \cdot  (  \bm{h}^{+}_{0,i} \times \bm{e}^{*}_{0,j} )+ \nabla_t \cdot  (  \bm{h}^{*}_{0,j} \times \bm{e}^{+}_{0,i} )\} dxdy$, $p_{ij}=\iint \{ \bm{z} \cdot ( \bm{e}^{*}_{0,j}  \times \bm{h}^{+}_{0,i})+ \bm{z} \cdot ( \bm{e}^{+}_{0,i}  \times \bm{h}^{*}_{0,j})\}dxdy $, $k_{ij}=ik_0\iint \Delta\epsilon(x,y)  \bm{e}^{*}_{0,j} \cdot \bm{e}^{+}_{0,i}  dxdy$. In this case, for $\Delta\epsilon(x,y)=0$, $b_{ij}+f_{ij} = i\left( {{\beta^* _{0,j}} - {\beta _{0,i}}} \right){p_{ij}}$ where $f_{ij}=-ik_0\iint \{(\bar{\bm{\mu}}_{r,0}+\bar{\bm{\mu}}_{r,0}^{T,*})h^*_j\cdot h^+_i+(\bar{\bm{\epsilon}}_{r,0}+\bar{\bm{\epsilon}}_{r,0}^{T,*})e^*_j\cdot e^+_i\}dxdy$. We will show in the next section that, CCMT works fine for Hermitian waveguides, but not for the case where non-Hermitian waveguides are considered.

\section{Results and Discussions}
\begin{figure}\centering
\includegraphics[scale=0.55]{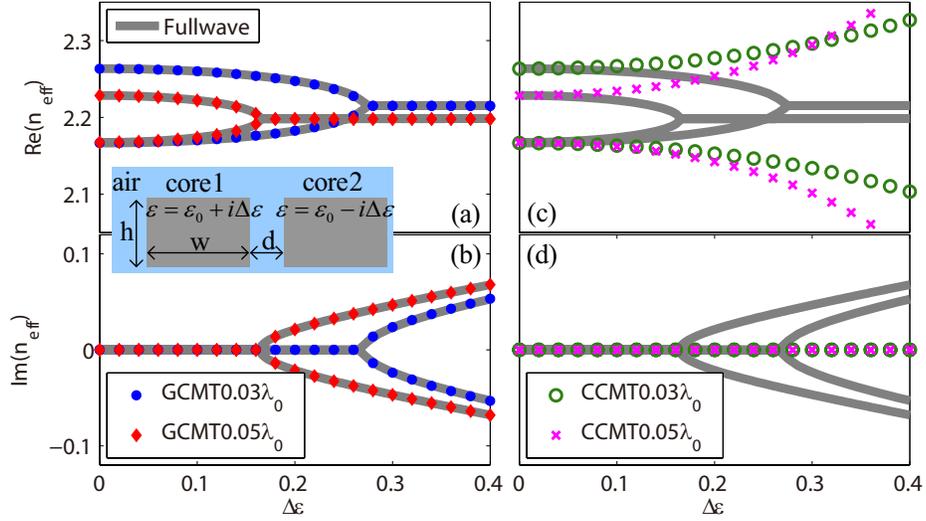}
\caption{\label{gainlossdisp} Real part and imaginary part of effective mode indices ($n_{eff}$) versus $\Delta\epsilon$ using GCMT (a,b) and CCMT (c,d). Gray solid lines are calculated from fullwave simulation. Inset shows the schematic diagram of two coupled core layers with loss and gain, surrounded by air. Dimensions are $h=0.2\lambda_0$, $w=0.3\lambda_0$, $d=0.03\lambda_0$ (blue solid circles, green open circles) or $0.05\lambda_0$ (red diamonds, magenta crosses), $\epsilon_0=10$. $\lambda_0$ is vacuum wavelength.}
\end{figure}

In the following, we use Eq. ~(\ref{gcmt}) to analyze dispersion relations in $\mathcal{PT}$-symmetric waveguides as discussed in \cite{EIganainyOL2007,PRL2008Moiseyev}. The structure of $\mathcal{PT}$-symmetric waveguides is shown by the inset of Fig.~\ref{gainlossdisp}. It is composed of two waveguides with identical geometry dimensions placed close  to each other. It is well known that the pair of  an even and odd super mode  is formed in this case. Phase transitions can be  observed as the magnitude of  the imaginary part of $\bar{\bm{\epsilon}}_r$ of two waveguides, i.e., $\bar{\bm{\epsilon}}_r=\bar{\bm{\epsilon}}_{r,0}+i\Delta\epsilon$ in core layer $1$ and $\bar{\bm{\epsilon}}_r=\bar{\bm{\epsilon}}_{r,0}-i\Delta\epsilon$ in core layer $2$, crosses a critical value as shown by the inset. This is used to create a symmetric index guiding profile and an anti-symmetric gain-loss profile. The gain/loss perturbation creates coupling between the odd and even mode pair so that the effective index of the two supermodes becomes closer and closer until an exceptional point where they become identical. Beyond the exceptional point, the real part of $n_{eff}$ remains the same, but the imaginary part of $n_{eff}$ of two modes break into two branches. When two waveguides are put more closer, the modes of two waveguides coupled more intensely so the even and odd supermodes have larger separation in the effective mode index. Therefore, it needs larger gain/loss parameter to get to the exceptional point. These are confirmed by COMSOL where two different gap size between two core layers are considered, as shown by the gray lines of Fig.~\ref{gainlossdisp}.

\begin{figure}\centering
\includegraphics[scale=0.5]{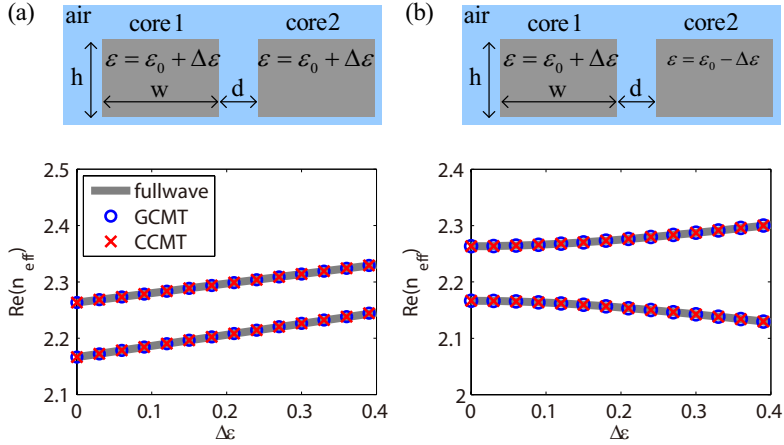}
\caption{\label{gainlossdisphermitianall} Real part of $n_{eff}$ versus $\Delta\epsilon$. Gray solid lines are calculated from fullwave simulations. Blue open circles (red crosses) represent results derived by GCMT(CCMT). $d=0.03\lambda_0$. Other parameters are the same as Fig. ~\ref{gainlossdisp}.}
\end{figure}

Then we apply our theory to predict the dispersion curves of the above mentioned structures. In this case, Eq.~(\ref{gcmt}) is an eigenvalue problem with the following form

\begin{equation}\label{matrixform}
\left[
\begin{array}{cc}
 \beta_{0,1}p_{11}-ik_{11}& \beta_{0,2}p_{12}-ik_{12}    \\
 \beta_{0,1}p_{21}-ik_{21}& \beta_{0,2}p_{22}-ik_{22}
\end{array}
\right]
\left[
\begin{array}{c}
 a_1    \\
 a_2
\end{array}
\right]=
\beta\left[
\begin{array}{cc}
 p_{11}& p_{12}    \\
 p_{21}& p_{22}
\end{array}
\right]
\left[
\begin{array}{c}
 a_1    \\
 a_2
\end{array}
\right]
\end{equation}
where $p_{ij}$ and $k_{ij}$ are defined in previous section. As a starting point, we use the mode fields provided by COMSOL at $\Delta\epsilon=0$. Propagation constants as well as eigenvectors are updated according to Eq.~(\ref{matrixform}) using mode fields at this point. Next, in-plane fields are recalculated using updated eigenvectors and used for deriving propagation constants as well as eigenvectors in the next step. By choosing a small step, full dispersion relations as a function of $\Delta\epsilon$ can be resolved. The solid symbols in Fig.~\ref{gainlossdisp}(a) and \ref{gainlossdisp}(b) show effective mode indices given by $n_{eff}=\beta/k_0$ derived by our method, where they match results from COMSOL remarkably well. For comparison, we also show the results obtained by CCMT in Fig.~\ref{gainlossdisp}(c) and \ref{gainlossdisp}(d). In this case \Eq{matrixform} becomes
\begin{equation}\label{matrixformconj}
\left[
\begin{array}{cc}
 \beta^*_{0,1}p_{11}-ik_{11}& \beta^*_{0,2}p_{12}-ik_{12}    \\
 \beta^*_{0,1}p_{21}-ik_{21}& \beta^*_{0,2}p_{22}-ik_{22}
\end{array}
\right]
\left[
\begin{array}{c}
 a_1    \\
 a_2
\end{array}
\right]=
\beta^*\left[
\begin{array}{cc}
 p_{11}& p_{12}    \\
 p_{21}& p_{22}
\end{array}
\right]
\left[
\begin{array}{c}
 a_1    \\
 a_2
\end{array}
\right].
\end{equation}
It is clear that in this case CCMT fails to capture the major feature of $\mathcal{PT}$-symmetric waveguides. However, instead of perturbing $\bar{\bm{\epsilon}}_r$ in the imaginary part, CCMT works fine for the case when $\bar{\bm{\epsilon}}_r$ is present the real part. Red crosses in Fig.  ~\ref{gainlossdisphermitianall}(a) shows the case that $\Delta\epsilon$ is real and increases with identical sign in both core layers. In this case, the separation between two mode indices remain the same but their absolute values increases as $\Delta\epsilon$ increases. Red crosses in Fig. ~\ref{gainlossdisphermitianall}(b) shows the case that $\Delta\epsilon$ is real but increases with opposite sign in two core layers. In this case, two mode indices are further separated as $\Delta\epsilon$ increases, indicating anti-crossing features. In both figures, only real part of $n_{eff}$ is shown since imaginary part of $n_{eff}$ in all cases is zero. Dispersion relations calculated according to GCMT are also shown in Fig. ~\ref{gainlossdisphermitianall} with blue open circles. Clearly, GCMT developed in this work gives the same results as CCMT does, agreeing well with full wave simulations given by gray lines shown in Fig. ~\ref{gainlossdisphermitianall}.

\section {Conclusions}
From reaction conservation, we provide a solid foundation for the construction of general couple mode theory that can handle mode hybridization in non-Hermitian waveguides. Using a scalar inner product, we establish the equivalence  between the self-adjointness of Maxwell's equations and reaction conservation. As for waveguide problems,  the dimension of the self-adjoint relation need to be reduced from 3D to 2D, in which the formula   turns out be non self-adjoint problem. Using coutering-propagating modes as the dual space of 2D non self-adjoint waveguide problem,  the eigenmodes  can be resolved from variational principles. Importantly, the  2D non self-adjoint relation can be elaborated into a set of coupled mode equation. We give a detailed  discussion of the  dimensional reduction for waveguide problem that relies on variational principle.  We then provide a procedure  of constructing GCMT for waveguide problem based on perturbation, which yields a set a coupled mode equations. To illustrate the effectiveness of GCMT developed in this work, it is applied to study the phase transition of coupled $\mathcal{PT}$-symmetric structures and shows excellent agreement with fullwave simulations. For comparison, results derived from CCMT are also shown, which fails to capture the major features of $\mathcal{PT}$-waveguides. Our theory provide direct analysis of eigenvalues of $\mathcal{PT}$-symmetric structures with in-cooperation of full vector fields. Thus it  might be useful to study and design non-Hermitian devices that support asymmetric and even nonreciprocal light propagation.

\section*{Acknowledgment}
This work was supported by National Natural Science Foundation of China (Grant No. 61405067 and 61405066) and Foundation for Innovative Research Groups of the Natural Science Foundation of Hubei Province (Grant No. 2014CFA004).

\end{document}